\documentstyle[seceq,preprint]{ptptex}
\newcommand{\bra}[1]{\langle {#1} |}     %%
\newcommand{\ket}[1]{| {#1} \rangle}     %%
\title{%        %You can use \\ for explicit line-break
Utility of $su(1,1)$-Algebra 
in a Schematic Nuclear $su(2)$-Model 
}
\author{%       %Use \sc for the family name
Hideaki {\sc Akaike}, Yasuhiko {\sc Tsue}, 
Constan\c{c}a {\sc Provid\^encia}$^{*}$, \\
Jo\~ao da {\sc Provid\^encia}$^{*}$, 
Atsushi {\sc Kuriyama}$^{**}$ and Masatoshi {\sc Yamamura}$^{**}$ 
}

\inst{%         %Affiliation, neglected when [addenda] or [errata]
Physics Division, Faculty of Science, Kochi University, Kochi 780-8520, 
Japan\\
$^{*}$Departamento de Fisica, Universidade de Coimbra, P-3000 Coimbra, 
Portugal\\
$^{**}$Faculty of Engineering, Kansai University, Suita 564-8680, Japan\\
}

\recdate{%      %Editorial Office will fill in this.
}

\abst{%       %this abstract is neglected when [addenda] or [errata]
The $su(2)$-algebraic model interacting with an environment 
is investigated from a viewpoint of treating the dissipative system. 
By using the time-dependent variational approach with a 
coherent state and with the help of the canonicity condition, 
the time-evolution of this quantum many-body system 
is described in terms of the canonical equations of motion in 
the classical mechanics. 
Then, it is shown that the $su(1,1)$-algebra plays an essential role 
to deal with this model. An exact solution with 
appropriate initial conditions is obtained by means of 
Jacobi's elliptic function. The implication to the dissipative 
process is discussed. 
}

\begin{document}

\maketitle

\section{Introduction}

One of the recent interests in nuclear theory is to 
investigate so-called thermal effect in nucleus and 
the dissipative process of the collective motion 
based on the quantum many-body theory or the quantum field theory.
To deal with the thermal effects in the quantum field theory, 
it is well known that 
there are some methods such as the imaginary-time formalism, 
the real-time formalism in thermo field dynamics and so on. 
Four of the present authors (Y.T., J.P., A.K. \& M.Y.) 
have proposed a method to describe the thermal effects 
and dissipation in many-body systems.\cite{TKY,KPTY} 
As for the description of the thermal effects in quantum 
many-particle systems, one of the key points in our formalism 
is to use the mixed-mode coherent state, which we have defined 
in the series of our papers,\cite{TKY,KPTY,KPTY2} 
as a trial state in the time-dependent variational approach. 
Furthermore, 
we learned that the $su(1,1)$-algebraic structure which 
the Hamiltonian has in the system under consideration 
plays an essential role when the non-equilibrium 
time-evolution and/or 
the dissipative process are realized 
in the quantum many-particle systems. 
Usually, so-called phase space doubling is carried out in order that 
the Hamiltonian has the $su(1,1)$-algebraic structure.\cite{V}
However, in general, 
the degree of freedom introduced by the phase space 
doubling cannot be interpreted as the environment or heat bath. 
We have given a formalism to describe the thermal effects 
and dissipation in a system in which the relevant and 
the irrelevant degrees of freedom interact each other and 
the irrelevant degree of freedom can be regarded as the 
environment.\cite{KPTY} 
Further, we have shown that the boson mapping theory\cite{MYT} 
presents the powerful technique to find the $su(1,1)$-algebraic 
behavior in the system governed by the $su(2)$-algebra. 
The recent review of the series of our work is seen 
in Ref.\citen{Suppl}

In this paper, we apply the above-mentioned formulation 
to the nuclear $su(2)$-model which interacts with the 
environment represented by a harmonic oscillator. 
Main purpose in this paper is to show the utility 
of the $su(1,1)$-algebra. As was mentioned previously, 
with the help of the $su(1,1)$-algebraic structure, 
we will find the possibility of the description of the 
dissipative process. We are restricted ourselves 
in this paper to the zero temperature system. 
Namely, we will adopt a 
coherent state as a trial state instead of the mixed-mode 
coherent state in the time-dependent variational principle. 
Further, 
the exact solution for the time-dependent variational 
equation of motion will be obtained with the help of the 
$su(1,1)$-algebra. It is shown that the solution is given 
in terms of the elliptic function. 
This solution is similar to that encountered in the 
investigation of nuclear collective motions in the 
classical $su(2)$-models.\cite{IY}
However, the physical situation is different.

This paper is organized as follows. In the next section, 
we introduce the nuclear $su(2)$-model interacting with a harmonic 
oscillator. The $su(1,1)$-algebraic structure is found 
by means of the Schwinger boson representation for 
the $su(2)$-algebra. The coherent state, which presents the 
classical counterpart of the original quantum many-body system, 
is introduced in \S 3. By the use of the canonicity conditions, 
the state is parameterized in terms of the canonical variables. 
Thus, it is shown that the equations of motion derived by 
the time-dependent variational principle are nothing but 
the canonical equations of motion. 
In \S 4, the equation of motion is solved in a certain case, 
and it is shown that the solution is expressed in terms of 
Jacobi's elliptic function. 
The implication to the dissipative process is discussed in \S 5. 
The last section is devoted to a concluding remarks.

\section{Nuclear $su(2)$-model and the construction of 
$su(1,1)$-algebra}

Let us start with the following Hamiltonian :
\begin{eqnarray}\label{2-1}
{\hat H}&=&\hbar\omega\left(a^*a+\frac{1}{2}\right)+2\epsilon{\hat S}_0 
-G{\hat S}_+{\hat S}_--\frac{\chi}{2}(
{\hat S}_+^2+{\hat S}_-^2)\nonumber\\
& &+\gamma\sqrt{\hbar}(a{\hat S}_++{\hat S}_- a^*) \ .
\end{eqnarray}
Here, $a^*$ and $a$ are boson creation and annihilation operators,
respectively, and the set $({\hat S}_0 , {\hat S}_\pm)$ composes 
the $su(2)$ algebra : 
\begin{eqnarray}\label{2-2}
& &[ a , a^*]=1 \ , \qquad\qquad [a , {\hat S}_{0,\pm}]=0 \ , \nonumber\\
& &[{\hat S}_+ , {\hat S}_-]=2\hbar {\hat S}_0 \ , \qquad
[{\hat S}_0 , {\hat S}_\pm]=\pm\hbar {\hat S}_\pm \ .
\end{eqnarray}
In the case $\omega=\chi=\gamma=0$, this model Hamiltonian 
is known as that of the pairing model in which the interaction 
is active for particle-particle pair with coupled angular 
momentum being 0. 
In the case $\omega=G=\gamma=0$, 
this corresponds to the Lipkin model which consists of two 
energy-levels with the same degeneracy. The level spacing is 
$2\hbar\epsilon$ and the interaction is active for particle-hole 
pair with coupled angular momentum being 0. 
These are the schematic 
nuclear models governed by the $su(2)$-algebra. We call these models 
the nuclear $su(2)$-models. In the case 
$G=\chi=0$, this model is nothing but the Jaynes-Cummings model 
if the representation of the $su(2)$-algebra is adopted as the spin 1/2 
one. The Jaynes-Cummings model 
is well known in the field of quantum optics. 
In general, we will regard in this paper the above model given 
in Eq.(\ref{2-1}) as a schematic nuclear model including 
both the relevant and the irrelevant degree of freedom which 
are represented by the $su(2)$-generators and $a$-boson, 
respectively. 
Namely, this model will be regarded as a typical model that 
the many-body system represented by the nuclear 
$su(2)$-model interacts with the environment represented by 
$a$-boson.
Other possibility to interpret this model 
is as follows : This model in Eq.(\ref{2-1}) 
can be interpreted as a schematic nuclear model including 
both the collective and the intrinsic degree of freedom which 
are represented by $a$-boson and the $su(2)$-generators, 
respectively.

In this paper, we use the Schwinger boson representation for the 
$su(2)$-generators.\cite{Schw} 
The new boson operators $(b, b^*)$ and 
$(c, c^*)$ are introduced and the $su(2)$-generators are 
expressed in terms of these two kinds of boson operators : 
\begin{equation}\label{2-3}
{\hat S}_0=\frac{\hbar}{2}(c^*c-b^*b) \ , \qquad
{\hat S}_+=\hbar c^* b \ , \qquad
{\hat S}_-=\hbar b^* c \ .
\end{equation}
The Casimir operator ${\hat \Gamma}_{su(2)}$ is written as
\begin{eqnarray}\label{2-4}
{\hat \Gamma}_{su(2)}&=&{\hat S}_0^2+\frac{1}{2}({\hat S}_+{\hat S}_- 
+{\hat S}_-{\hat S}_+)={\hat S}({\hat S}+\hbar) \ , \\
{\hat S}&=&\frac{\hbar}{2}(c^*c+b^*b) \ . \nonumber
\end{eqnarray}

Further, the following operators can be also defined in terms of 
the original boson operators $(a, a^*)$ and the Schwinger boson 
$(b, b^*)$ as 
\begin{equation}\label{2-5}
{\hat T}_0=\frac{\hbar}{2}(bb^*+a^*a) \ , \qquad
{\hat T}_+=\hbar b^* a^* \ , \qquad
{\hat T}_-=\hbar a b \ .
\end{equation}
These operators compose the $su(1,1)$-algebra, and the 
Casimir operator ${\hat \Gamma}_{su(1,1)}$ is also defined as 
follows : 
\begin{eqnarray}
& &[{\hat T}_+ , {\hat T}_-]=-2\hbar{\hat T}_0 \ ,\qquad
[{\hat T}_0 , {\hat T}_\pm ]=\pm \hbar{\hat T}_\pm \ , 
\label{2-6}\\
& &{\hat \Gamma}_{su(1,1)}={\hat T}_0^2
-\frac{1}{2}({\hat T}_+{\hat T}_-+{\hat T}_-{\hat T}_+)
={\hat T}({\hat T}-\hbar) \ , \label{2-7}\\
& &\ \ {\hat T}=\frac{\hbar}{2}(bb^*-a^*a) \ .\nonumber
\end{eqnarray}
It should be noted that 
the two algebras are not independent in this model. 
Actually, we can calculate the commutation relations : 
$[{\hat S} , {\hat T}_0]=[{\hat S} , {\hat T}]=
[{\hat T} , {\hat S}_0]=0$. However, we obtain 
$[{\hat S}, {\hat T}_\pm]\neq 0$ and $[{\hat T}, {\hat S}_\pm]\neq 0$.

The Hamiltonian (\ref{2-1}) is then expressed by using 
the $su(2)$- and the $su(1,1)$-generators together with 
their Casimir operators. 
The Hamiltonian is recast into 
\begin{eqnarray}
{\hat H}&=&{\hat H}_0+{\hat H}_1+{\hat H}_2 \ , \label{2-8}\\
{\hat H}_0&=&\hbar\omega\left(a^*a+\frac{1}{2}\right)+
2\epsilon{\hat S}_0-G{\hat S}_+{\hat S}_- \nonumber\\
&=&\omega\left({\hat T}_0-{\hat T}+\frac{\hbar}{2}\right)
\nonumber\\
& &+2\epsilon({\hat S}-{\hat T}_0-{\hat T}+\hbar)
-G({\hat T}_0+{\hat T})(2{\hat S}-{\hat T}_0-{\hat T}+\hbar) 
\ , \label{2-9}\\
{\hat H}_1&=&\gamma\sqrt{\hbar}(a{\hat S}_++{\hat S}_-a^*)
=\gamma\sqrt{\hbar}({\hat T}_-c^*+c{\hat T}_+) \ , 
\label{2-10}\\
{\hat H}_2&=&-\frac{\chi}{2}({\hat S}_+^2+{\hat S}_-^2) \ .
\label{2-11}
\end{eqnarray}
In any case, we have the commutation 
relation, $[{\hat H}, {\hat S}]=0$. Thus, ${\hat S}$ is 
a constant of motion. Besides ${\hat S}$, we have conserved 
variables in some cases : 
(i) If $\chi=0$, that is, the case of no particle-hole 
interaction, then ${\hat H}$ does not contain 
${\hat S}_\pm$. Thus, we can derive $[{\hat H}, {\hat T}]=0$, so 
that ${\hat T}$ is a constant of motion in addition to 
${\hat S}$.  (ii) If $\gamma=0$, that is, there is no 
interaction between the relevant and the irrelevant motion, 
then ${\hat H}$ does 
not contain any linear term with respect to $a$ and $a^*$. 
Thus, we obtain $[{\hat H}, {\hat T}_0-{\hat T}]=0$ because 
${\hat T}_0-{\hat T}=\hbar a^*a$. Namely, we have another 
constant of motion ${\hat T}_0-{\hat T}$ in addition to ${\hat S}$. 
(iii) If $\chi=\gamma=0$, then $[{\hat H}, {\hat T}]=[{\hat H}, 
{\hat T}_0]=0$. 
Thus the variables ${\hat T}$ and ${\hat T}_0$ are conserved 
independently.

In any case, the operators 
${\hat S}$, ${\hat T}$, ${\hat T}_0$ and ${\hat S}_0$
commute each other. However, ${\hat S}_0$ is expressed in 
terms of the other three operators as 
${\hat S}_0={\hat S}-{\hat T}_0-{\hat T}+\hbar$. 
Thus, it is enough to introduce the simultaneous eigenstate, 
$\ket{s,t,t_0}$, for ${\hat S}$, ${\hat T}$ and ${\hat T}_0$ 
with the eigenvalues $\hbar s$, $\hbar t$ and $\hbar t_0$, 
respectively. 
Except for the normalization constant, the eigenstate 
is given as 
\begin{equation}\label{2-12}
\ket{s,t,t_0}=(a^*)^{t_0-t}(b^*)^{t_0+t-1}(c^*)^{2s-t-t_0+1}\ket{0} \ .
\end{equation}
Here, $a\ket{0}=b\ket{0}=c\ket{0}=0$. The eigenvalue equations 
are written as 
\begin{eqnarray}\label{2-13}
& &{\hat S}\ket{s,t,t_0}=\hbar s\ket{s,t,t_0} \ , \quad
{\hat T}\ket{s,t,t_0}=\hbar t\ket{s,t,t_0} \ , \quad
{\hat T}_0\ket{s,t,t_0}=\hbar t_0\ket{s,t,t_0} \ , \nonumber\\
& &{\hat S}_0\ket{s,t,t_0}=\hbar (s-t-t_0+1)\ket{s,t,t_0} \ .
%\nonumber
\end{eqnarray}
As for the $su(1,1)$-algebra, the case $t \ge 1/2$ will only 
be treated from now on. In this case, the eigenstate 
$\ket{s,t,t_0}$ is expressed by the use of ${\hat S}_+$ and 
${\hat T}_+$ as 
\begin{equation}\label{2-14}
\ket{s,t,t_0}={\cal N}({\hat T}_+)^{t_0-t}
({\hat S}_+)^{2s-t-t_0+1}(b^*)^{2s-t_0+t}\ket{0} \ .
\end{equation}
Then, we can show the following relations :
\begin{equation}\label{2-15}
{\hat S}_- (b^*)^{2s-t_0+t}\ket{0}={\hat T}_-(b^*)^{2s-t_0+t}\ket{0}
=0 \ .
\end{equation}
Therefore, ${\hat S}_+$ and ${\hat T}_+$ play a role of the raising 
operators from the state $(b^*)^{2s-t_0+t}\ket{0}$.

\section{Construction of coherent state and classical counterpart}

In the previous section, we have introduced the eigenstate 
with respect to the operators ${\hat S}$, ${\hat T}$, 
${\hat T}_0$ and ${\hat S}_0$. Then, ${\hat S}_+$ and ${\hat T}_+$ 
are regarded as the raising operators. Thus, it is allowed 
to construct the following coherent state : 
\begin{equation}\label{3-1}
\ket{c}=N_c \exp\left(VU^{-1}/\hbar\cdot {\hat T}_+\right)
\cdot\exp\left(UYW^{-1}/\hbar\cdot{\hat S}_+\right)
\cdot \exp\left(\sqrt{2/\hbar}\cdot WU^{-1}b^*\right)
\ket{0} \ .
\end{equation}
Here, $V$, $W$ and $Y$ are complex variables, 
$U=\sqrt{1+|V|^2}$ and $N_c$ is a normalization factor. 
The state $\ket{c}$ is rewritten as 
\begin{eqnarray}\label{3-2}
\ket{c}&=&
N_c\exp\left(\sqrt{\frac{2}{\hbar}}Yc^*+
VU^{-1}b^*a^*+\sqrt{\frac{2}{\hbar}}WU^{-1}b^*\right)\ket{0}
\nonumber\\
&=&\ket{c_c}\otimes \ket{c_{ab}} \ , \\
& &\ket{c_c}=e^{-\frac{1}{\hbar}|Y|^2}\cdot\exp\left(
\sqrt{\frac{2}{\hbar}}Yc^*\right)\ket{0} \ , \nonumber\\
& &\ket{c_{ab}}=U^{-1}e^{-\frac{1}{\hbar}|W|^2}
\cdot\exp\left(VU^{-1}b^*a^*+\sqrt{\frac{2}{\hbar}}WU^{-1}b^*
\right)\ket{0} \ .\nonumber
\end{eqnarray}
By introducing the following new operators, 
\begin{equation}\label{3-4}
a'=Ua-Vb^* \ , \qquad
b'=Ub-Va^*-\sqrt{2/\hbar}\ W \ ,\quad
c'=c-\sqrt{2/\hbar}\ Y \ , 
\end{equation}
it is shown that the state $\ket{c}$ is vacuum for 
$a'$, $b'$ and $c'$ :
\begin{equation}\label{3-5}
a'\ket{c}=b'\ket{c}=c'\ket{c}=0 \ .
\end{equation}
Inversely, $a$, $b$ and $c$ are expressed as 
\begin{equation}\label{3-6}
a=Ua'+Vb'^*+\sqrt{2/\hbar}\ W^*V \ , \quad
b=Va'^*+Ub'+\sqrt{2/\hbar}\ WU\ ,\quad
c=c'+\sqrt{2/\hbar}\ Y \ .
\end{equation}

By the use of the above relations, we can easily calculate 
the expectation values of the $su(2)$- and the $su(1,1)$-generators 
and their Casimir operators. For example, 
\begin{eqnarray}\label{3-7}
& &T=\bra{c}{\hat T}\ket{c}=|W|^2+\frac{\hbar}{2} \ , \nonumber\\
& &S=\bra{c}{\hat S}\ket{c}=\left(|W|^2+\frac{\hbar}{2}\right)
|V|^2+|W|^2+|Y|^2 \ , \nonumber\\
& &T_0=\bra{c}{\hat T}_0\ket{c}=\left(|W|^2+\frac{\hbar}{2}\right)
\left(1+2|V|^2\right) \ .
\end{eqnarray}

It is here noted that the dynamical variables are 
$(V, V^*, W, W^*, Y, Y^*)$. 
As is similar to the time-dependent Hartree-Fock theory with canonical 
form,\cite{YK87} 
it is convenient to parameterize the coherent state 
in terms of canonical variables. For this purpose, 
we impose the following canonicity conditions : 
\begin{eqnarray}\label{3-8}
& &\bra{c}i\hbar\partial_{\phi_0}\ket{c}=T_0-\frac{\hbar}{2} \ , 
\qquad
\bra{c}i\hbar\partial_{T_0}\ket{c}=0 \ , \nonumber\\
& &\bra{c}i\hbar\partial_{\phi}\ket{c}=T-\frac{\hbar}{2} \ , 
\qquad\ \ 
\bra{c}i\hbar\partial_{T}\ket{c}=0 \ , \nonumber\\
& &\bra{c}i\hbar\partial_{\psi}\ket{c}=S \ , 
\qquad\qquad\ 
\bra{c}i\hbar\partial_{S}\ket{c}=0 \ .
\end{eqnarray}
The sets of variables $(T_0,\phi_0),\ (T, \phi)$ and $(S, \psi)$ 
are those of the canonical variables. 
The above canonicity conditions can be easily solved by the 
use of the following relation : 
\begin{eqnarray}\label{3-9}
\bra{c}i\hbar\partial_z\ket{c}
&=&i\Bigl[
(|W|^2+\hbar/2)\cdot(V^*\partial_z V-V\partial_z V^*) \nonumber\\
& &+(W^*\partial_z W-W\partial_z W^*)
+(Y^*\partial_z Y-Y\partial_z Y^*)\Bigl] \ .
\end{eqnarray}
As a result, the original variables are expressed by 
the canonical variables as 
\begin{eqnarray}\label{3-10}
& &V=\sqrt{\frac{T_0-T}{2T}}\cdot
e^{-i\phi_0}e^{-i\psi/2} \ , \nonumber\\
& &W=\sqrt{T-\frac{\hbar}{2}}\cdot e^{-i\phi_0/2}e^{-i\phi/2}
e^{-i\psi/2} \ , \nonumber\\
& &Y=\sqrt{\frac{1}{2}(2S-T_0-T+\hbar)}\cdot e^{-i\psi/2} \ .
\end{eqnarray}
Thus, we can obtain the expectation values of various 
operators with respect to $\ket{c}$ 
in terms of the canonical variables :
\begin{eqnarray}\label{3-11}
& &\bra{c}{\hat S}_0\ket{c}=S-T_0-T+\hbar \ , \nonumber\\
& &\bra{c}{\hat S}_+\ket{c}=\sqrt{2(2S-T_0-T+\hbar)}
\sqrt{T-\frac{\hbar}{2}}\sqrt{\frac{T_0+T}{2T}}
\cdot e^{-i(\phi_0+\phi)/2} \ , \nonumber\\
& &\bra{c}{\hat T}_0\ket{c}=T_0 \ , \nonumber\\
& &\bra{c}{\hat T}_+\ket{c}=\sqrt{(T_0-T)(T_0+T)}\cdot
e^{i\phi_0}e^{i\psi/2} \ , \nonumber\\
& &\bra{c}{\hat S}_+{\hat S}_-\ket{c}
=(2S-T_0-T+\hbar)(T_0+T) \ , \nonumber\\
& &\bra{c}a^*a\ket{c}=(T_0-T)/\hbar \ , \nonumber\\
& &\bra{c}{\hat S}_+^2\ket{c}
=\left(1-\frac{\hbar}{2T}\right)(T_0+T)(2S-T_0-T+\hbar)
e^{-i(\phi_0+\phi)} \ , \nonumber\\
& &\bra{c}{\hat S}_-^2\ket{c}
=\left(1-\frac{\hbar}{2T}\right)(T_0+T)(2S-T_0-T+\hbar)
e^{i(\phi_0+\phi)} \ , \nonumber\\
& &\bra{c}\sqrt{\hbar}a^*{\hat S}_-\ket{c}
=\sqrt{(T_0-T)(T_0+T)(2S-T_0-T+\hbar)}\ 
e^{i\phi_0} \ , \nonumber\\
& &\bra{c}\sqrt{\hbar}a{\hat S}_+\ket{c}
=\sqrt{(T_0-T)(T_0+T)(2S-T_0-T+\hbar)}\ 
e^{-i\phi_0} \ .
\end{eqnarray}
Thus, the expectation value of the Hamiltonian with respect 
to $\ket{c}$ is given by 
\begin{eqnarray}\label{3-12}
H=\bra{c}{\hat H}\ket{c}&=&H_0+H_1+H_2 \nonumber\\
H_0&=&\bra{c}{\hat H}_0\ket{c}=\omega(T_0-T+\hbar/2)+2\epsilon(S-T_0-T+\hbar)
\nonumber\\
& &\qquad\qquad\quad
-G(2S-T_0-T+\hbar)(T_0+T) \nonumber\\
& &\qquad\qquad 
=f(T_0, T ;S) \ ,\nonumber\\
H_1&=&\bra{c}{\hat H}_1\ket{c} 
=2\gamma\sqrt{(T_0-T)(T_0+T)(2S-T_0-T+\hbar)}\ 
\cos\phi_0 \nonumber\\
& &\qquad\qquad 
=g(T_0, T ;S)\cos\phi_0\ ,  \nonumber\\
H_2&=&\bra{c}{\hat H}_2\ket{c} 
=-\chi(1-\hbar/(2T))(T_0+T)(2S-T_0-T+\hbar)\ 
\cos(\phi_0+\phi) \nonumber\\
& &\qquad\qquad 
=h(T_0, T ;S)\cos(\phi_0+\phi) \ .
\end{eqnarray}
Here, ${\hat H}_0$, ${\hat H}_1$ and ${\hat H}_2$ have been 
defined in Eqs.(\ref{2-9})$\sim$(\ref{2-11}).

The equations of motion are derived from the time-dependent 
variational principle :
\begin{equation}\label{3-13}
\delta\int_{t_1}^{t_2} dt \bra{c} i\hbar\partial_t -{\hat H} 
\ket{c}=0 \ .
\end{equation}
With the help of the canonicity conditions (\ref{3-8}), 
the derived equations of motion have the form of 
canonical equations of motion : 
\begin{eqnarray}
& &{\dot T}_0=-\frac{\partial H}{\partial \phi_0}
=g(T_0, T ;S)\sin\phi_0 + h(T_0, T; S)\sin(\phi_0+\phi) \ ,\nonumber\\
& &{\dot \phi}_0=\frac{\partial H}{\partial T_0}\nonumber\\
& &\quad\ =\frac{\partial f(T_0,T;S)}{\partial T_0}
+\frac{\partial g(T_0,T;S)}{\partial T_0}\cos\phi_0
+\frac{\partial h(T_0,T;S)}{\partial T_0}\cos(\phi_0+\phi) 
\ , \quad\label{3-14}\\
& &{\dot T}=-\frac{\partial H}{\partial \phi}
=h(T_0, T ;S)\sin(\phi_0+\phi) \ ,\nonumber\\
& &{\dot \phi}=\frac{\partial H}{\partial T}\nonumber\\
& &\quad =\frac{\partial f(T_0,T;S)}{\partial T}
+\frac{\partial g(T_0,T;S)}{\partial T}\cos\phi_0
+\frac{\partial h(T_0,T;S)}{\partial T}\cos(\phi_0+\phi) 
\ , \quad\label{3-15}\\
& &{\dot S}=-\frac{\partial H}{\partial \psi}
=0 \ ,\nonumber\\
& &{\dot \psi}=\frac{\partial H}{\partial S}\nonumber\\
& &\quad\ =\frac{\partial f(T_0,T;S)}{\partial S}
+\frac{\partial g(T_0,T;S)}{\partial S}\cos\phi_0
+\frac{\partial h(T_0,T;S)}{\partial S}\cos(\phi_0+\phi) 
\ . \quad\label{3-16}
\end{eqnarray}
Let us consider simple cases : 
(i) If $\chi=0$, then $h(T_0,T; S)=0$. In this case, 
from Eq.(\ref{3-15}), the quantity $T$ is conserved in 
addition to $S$. 
This fact is, of course, originated from the fact 
$[{\hat H}, {\hat T}]=0$ mentioned in \S 2. 
It is thus necessary to solve the time-evolution 
of $T_0$ : 
\begin{eqnarray}\label{3-17}
& &{\dot T}_0=g(T_0, T; S)\sin\phi_0 \ , \nonumber\\
& &{\dot \phi}_0=\frac{\partial f(T_0,T;S)}{\partial T_0}
+\frac{\partial g(T_0,T;S)}{\partial T_0}\cos \phi_0 \ .
\end{eqnarray}
The expectation value of the total Hamiltonian, which is 
a constant of motion, is expressed  as
\begin{equation}\label{3-18}
H=f(T_0, T; S)+g(T_0, T; S)\cos\phi_0=E_0 \ .
\end{equation}
From Eqs.(\ref{3-18}) and (\ref{3-17}), we obtain 
\begin{equation}\label{3-19}
\left(\frac{{\dot T}_0}{g(T_0, T; S)}\right)^2
+\left(\frac{E_0-f(T_0, T; S)}{g(T_0, T; S)}\right)^2=1 \ .
\end{equation}
Namely, 
\begin{equation}\label{3-20}
{\dot T}_0^2=[g(T_0, T; S)]^2-[E_0-f(T_0, T; S)]^2 \ .
\end{equation}
(ii) If $\gamma=0$, then $g(T_0,T; S)=0$. In this case, 
from Eqs.(\ref{3-14}) and (\ref{3-15}), the quantity 
$T_0-T$ is conserved in 
addition to $S$. It is thus necessary to solve the time-evolution 
of $T_0-T$. However it is enough to know the time-dependence 
for $T_0$ because ${\dot T_0}-{\dot T}=0$ : 
\begin{eqnarray}\label{3-21}
& &{\dot T}_0=h(T_0, T; S)\sin(\phi_0+\phi) \ , \nonumber\\
& &{\dot \phi}_0=\frac{\partial f(T_0,T;S)}{\partial T_0}
+\frac{\partial h(T_0,T;S)}{\partial T_0}\cos (\phi_0+\phi) \ .
\end{eqnarray}
The expectation value of the total Hamiltonian is expressed  as
\begin{equation}\label{3-22}
H=f(T_0, T; S)+h(T_0, T; S)\cos(\phi_0+\phi)=E_0 \ .
\end{equation}
From Eqs.(\ref{3-21}) and (\ref{3-22}), we obtain 
\begin{equation}\label{3-23}
\left(\frac{{\dot T}_0}{h(T_0, T; S)}\right)^2
+\left(\frac{E_0-f(T_0, T; S)}{h(T_0, T; S)}\right)^2=1 \ .
\end{equation}
Namely, 
\begin{equation}\label{3-24}
{\dot T}_0^2=[h(T_0, T; S)]^2-[E_0-f(T_0, T; S)]^2 \ .
\end{equation}
Our next task is to solve the equation of motion (\ref{3-20}) 
or (\ref{3-24}) which have the same form.

\section{Utility of elliptic function for a 
solution of equations of motion}

In the previous section, it has been shown that in the simple cases 
the equation of motion for $T_0$ which has 
a form in Eq.(\ref{3-20}) or (\ref{3-24}) should be solved. 
Hereafter, let us consider the case (i), that is, $\chi=0$, 
in which the particle-hole interaction is not active. 
In the region $T_0\ge T\ge \hbar/2$ and the adequate model parameters, 
the right-hand side of Eq.(\ref{3-20}) is 
always non-negative. However, we regard the right-hand side 
of Eq.(\ref{3-20}) as a function of $T_0$ with arbitrary parameter 
set : 
\begin{eqnarray}\label{4-1}
{\dot T}_0^2=F(T_0)&=&
[g(T_0, T; S)]^2-[E-f(T_0, T; S)]^2 \nonumber\\
&=&-G^2(T_0-\tau_0)(T_0-\tau_1)(T_0-\tau_2)(T_0-\tau_3) \ ,
\end{eqnarray}
where we used the knowledge that the function $F$ is the 
function of fourth order for $T_0$ and the coefficient of 
$T_0^4$ is negative : $F(T_0)=-G^2 T_0^4+\cdots$. 
It should be noted here 
that, in case (ii), if the parameters satisfy 
an inequality $\chi^2(1-\hbar/(2T))^2<G^2$, the same situation 
as the case (i) is realized. 
From the fact $F(T_0\rightarrow\pm\infty)<0$, it is concluded that 
the two real solutions for $F(T_0)=0$ except for 
${\dot T}_0\equiv 0$ must exist because 
the physical region $F>0$ must exist. 
In addition to this fact, in the second line in Eq.(\ref{4-1}), 
we assume that there exist four real solutions for $F(T_0)=0$ 
which are denoted as $\tau_0$, $\tau_1$, $\tau_2$ and $\tau_3$ 
with $\tau_0 < \tau_1 < \tau_2 < \tau_3$. 
We can numerically check the existence of four real solutions 
for $F(T_0)=0$. For example, if the parameters are 
adopted as $\hbar G/\epsilon=1/2$, $\omega/\epsilon=1$, 
$\gamma\sqrt{\hbar}/\epsilon=0.1$ and $E_0/\epsilon=0$ and 
the variables are $S/\hbar=5$ and $T/\hbar=1/2$, the above-mentioned 
situation is realized. Further, in this parameter set, the physical 
region is given in $\tau_0 < T_0 < \tau_1$. 
Thus, hereafter, the physical region 
is taken as $\tau_0 <T_0 < \tau_1$. 
Then, $T_0$ corresponding to the physical value is parameterized as follows :
\begin{equation}\label{4-2}
T_0=\tau_0-(\tau_0-\tau_1)x^2\ , \qquad -1\le x \le 1 \ .
\end{equation}
Thus, we can derive the equation for $x$ from Eq.(\ref{4-1}) : 
\begin{equation}\label{4-3}
{\dot x}^2=\frac{G^2}{4}(\tau_0-\tau_2)(\tau_0-\tau_3)
(1-x^2)(1-k_1 x^2)(1-k_2x^2) \ , 
\end{equation}
where $k_1$ and $k_2$ are defined as 
\begin{equation}\label{4-4}
k_1=\frac{\tau_0-\tau_1}{\tau_0-\tau_2} \ , \quad
k_2=\frac{\tau_0-\tau_1}{\tau_0-\tau_3} \ , \quad
0\le k_2 \le k_1 \le 1 \ .
\end{equation}
Further, if we set 
\begin{equation}\label{4-5}
x(t)=\frac{m(t)}{\sqrt{1-k_2+k_2m^2(t)}} \ ,\qquad (-1\le m \le 1)
\end{equation}
the variable $m(t)$ satisfies the following differential 
equation : 
\begin{eqnarray}
& &{\dot m}^2=Z^2 (1-m^2)(1-k^2m^2) \ ,
\label{4-6}\\
& &Z=\frac{K}{\pi}\omega_0\ , \label{4-7}
\end{eqnarray}
where $\omega_0$, $K$ and $k$ are defined as 
\begin{eqnarray}\label{4-8}
& &\omega_0=\frac{G}{2}
\sqrt{(\tau_0-\tau_2)(\tau_1-\tau_3)}\ \frac{\pi}{K} \ , 
\nonumber\\
& &K=\int_0^1\frac{dx}{\sqrt{(1-x^2)(1-k^2x^2)}}=K(k) \ , 
\nonumber\\
& &k=\sqrt{\frac{k_1-k_2}{1-k_2}}
=\sqrt{\frac{(\tau_0-\tau_1)(\tau_2-\tau_3)}
{(\tau_0-\tau_2)(\tau_1-\tau_3)}} \ . \qquad
(0\le k \le 1)
\end{eqnarray}
Here, $K$ and $k$ are the complete elliptic integral of the first kind 
and the modulus, respectively.
If we take the initial value $m(t=0)=0$, that is, $T_0(t=0)=\tau_0$, 
we thus obtain the following solution for $m(t)$ : 
\begin{equation}\label{4-9}
m(t)={\rm sn} (Zt, k) \ , 
\end{equation}
where ${\rm sn} (Zt, k)$ is Jacobi's elliptic function :
\begin{equation}\label{4-10}
Zt=\int_0^{m(t)} \frac{dm}{\sqrt{(1-m^2)(1-k^2m^2)}} \ .
\end{equation}
As a result, we obtain $T_0$ as a function of time $t$ :
\begin{equation}\label{4-11}
T_0=\tau_0-\frac{(\tau_0-\tau_1)(\tau_0-\tau_3){\rm sn}^2(Zt, k)}
{(\tau_1-\tau_3)+(\tau_0-\tau_1){\rm sn}^2(Zt, k)} \ . 
\end{equation}
It should be noted that, if the physical region is given by 
$\tau_2< T_0 < \tau_3$, the solution for $T_0$ is also 
expressed in terms of Jacobi's elliptic function 
similar to Eq.(\ref{4-11}).

In the Fourier series representation for Jacobi's elliptic 
function, the function $k^2{\rm sn}^2(Zt,k)$ is 
expressed as 
\begin{equation}\label{4-12}
k^2{\rm sn}^2(Zt, k)=1-\frac{E}{K(k)}-\frac{\pi^2}{K(k)^2}
\cdot\sum_{n=1}^\infty \frac{n}{\sinh (n\pi K(k')/K(k))}
\cdot\cos(n\omega_0t) \ ,
\end{equation}
where $k'=\sqrt{1-k^2}$ and $E$ represents 
the complete elliptic integral of the second kind :
\begin{equation}\label{4-13}
E=\int_0^1 \sqrt{\frac{1-k^2x^2}{1-x^2}}dx \ .
\end{equation}
From Eqs.(\ref{4-11}) and (\ref{4-12}), it can be seen that 
the time-dependent variable $T_0$ has a fundamental period 
$t_0$ given by 
\begin{equation}\label{4-14}
t_0=\frac{2\pi}{\omega_0}\ .
\end{equation}

\section{Discussion}

As was mentioned in \S 1, it is interesting to investigate 
the dissipative process which is governed by the $su(1,1)$-algebraic 
structure. In this model-Hamiltonian we are treating in this paper, 
the $su(1,1)$-algebraic structure is hidden and appears by means of 
the Schwinger boson representation for $su(2)$-generators. 
In the previous section, the parameter $\chi$, which is the strength 
of the particle-hole interaction, is taken as 0. 
It has been shown that in this case there exists a fundamental 
period $t_0$ in the solution expressed by the elliptic function. 
However, if this period is long compared with the typical time scale 
we consider, the dissipative situation may be realized. 
In this section, we will investigate this possibility. 

The fundamental period $t_0$ is given in Eq.(\ref{4-14}). 
Therefore, if the angular frequency $\omega_0$, which 
has been given in Eq.(\ref{4-8}), is infinitesimal or 
very small value, the period $t_0$ substantially becomes infinity. 
Thus, the periodicity disappears. In order to realize this 
situation, it is necessary that the modulus $k$ is close to 1 
because the complete elliptic integral of the first kind, $K$, 
reveals the behavior of 
a logarithmic divergence with $k\rightarrow 1$. 
As a result, $\omega_0$ is close to 0 and $t_0$ becomes infinity. 
If $\tau_1\sim\tau_2$, then $k\sim 1$ is realized. 
In this case, Jacobi's elliptic function ${\rm sn} (Zt, k=1)$ is 
expressed in terms of the hyperbolic tangent, namely, 
\begin{equation}\label{5-1}
{\rm sn}(Zt, 1)=\tanh Zt \ .
\end{equation}
Thus, the solution $T_0$ is expressed by 
\begin{equation}\label{5-2}
T_0=\tau_0-\frac{\tau_0-\tau_3}{\frac{\tau_1-\tau_3}{\tau_0-\tau_1}
\cdot\frac{1}{\tanh^2 Zt}+1} \ .
\end{equation}
This is a monotonically increasing function from 
$\tau_0$ at $t=0$ to $\tau_1$ at $t\rightarrow\infty$. 
Thus, the periodicity is lost. 

In this special case, the energy of the degree of freedom 
represented by $a$-boson increases monotonically, that is, 
\begin{equation}\label{5-3}
E_a=\bra{c}\hbar\omega(a^*a+1/2)\ket{c}=\omega\left(T_0-T+\frac{\hbar}{2}
\right)
\end{equation}
increases for time because $T_0$ increases monotonically. 
Here, it is possible to introduce a ``coordinate" $x$ 
which is defined 
and expressed in terms of the canonical variables :
\begin{eqnarray}\label{5-4}
{\hat x}&=&\sqrt{\frac{\hbar}{2m\omega}}(a+a^*) \ , \nonumber\\
x&=&\bra{c}{\hat x}\ket{c}=\sqrt{\frac{\hbar}{2m\omega}}
\sqrt{\frac{2}{\hbar}}(W^*V+WV^*) \nonumber\\
&=&\frac{2}{\sqrt{m\omega}}\cdot\sqrt{T-\frac{\hbar}{2}}
\sqrt{\frac{T_0-T}{2T}}\ \cos\frac{\phi-\phi_0}{2} \ .
\end{eqnarray}
It is seen that the amplitude of $x$ increases monotonically. 
Thus, $x$ reveals the behavior of the amplified oscillation. 
If the degrees of freedom represented by $a$-boson and by 
the $su(2)$-generators are regarded as 
the environment and the relevant degree 
of freedom, respectively, the process in which 
the energy of the relevant motion dissipates to the environment 
is described. 
This is nothing but the dissipative process. 
This process may be interpreted as the 
``cooling" process for the relevant motion, 
although we do not treat the system at finite 
temperature. 
Actually, if the state $\ket{c}$ is replaced into the 
mixed-mode coherent state, the thermal effect may be treated 
in this framework.

\section{Concluding remarks}

We have investigated the nuclear $su(2)$-model interacting with 
the harmonic oscillator. By the use of the Schwinger boson 
representation for the $su(2)$-algebra, the $su(1,1)$-algebraic 
behavior was come in sight. 
In this model we have considered in this paper, 
it is allowed to obtain 
the exact solution for the classical equation of motion. 
It has been shown that the solution is expressed in terms of 
Jacobi's elliptic function. The solution has a fundamental 
period. However, if the period is so long, it may be possible 
to realize the dissipative process. 
In this paper, we are restricted ourselves to deal with the 
zero temperature case. If we apply our formalism to the finite 
temperature system, it is necessary to replace the coherent state 
into the mixed-mode coherent state. We have already investigated 
the mixed-mode coherent state in the many-body system 
composed of three kinds of boson operators.\cite{KPTY6} 
The investigation of the behavior in the nuclear $su(2)$-model 
interacting with the environment at finite temperature is 
left as a future problem.

\section*{Acknowledgements}

Main part of this work was completed when two of the present authors 
(Y. T. \& M. Y.) stayed at Coimbra in September, 2000. 
They express their sincere thanks to Professor 
J. da Provid\^encia, co-author of this paper, for his kind invitation 
to Coimbra.

\end{document}